\documentclass[pra,aps,eqsecnum,amsmath,amssymb,showpacs,showkeys]{revtex4}
\usepackage{xtheorem}
\newcommand{\am}{\mathsf{A}}
\newcommand{\bm}{\mathsf{B}}
\newcommand{\ax}{\mathsf{X}}
\newcommand{\ay}{\mathsf{Y}}
\newcommand{\az}{\mathsf{Z}}
\newcommand{\dm}{\mathsf{D}}
\newcommand{\pen}{\openone}
\newcommand{\hh}{\mathcal{H}}

\newcommand{\tr}{{\rm{Tr}}}
\newcommand{\rd}{{\rm{D}}}
\newcommand{\kr}{{\rm{ker}}}
\newcommand{\sop}{{\rm{supp}}}
\newcommand{\spc}{{\rm{spec}}}

\newtheorem{plain}{Thm}{Theorem}[section]
{Lemma}
{Definition}
{Remark}

\begin{document}
\clearpage
\preprint{}

\title{{\bf Upper continuity bounds on the relative $q$-entropy for $q>1$}}

\author{Alexey E. Rastegin}
 \affiliation{Department of Theoretical Physics, Irkutsk State University,
Gagarin Bv. 20, Irkutsk 664003, Russia}

\begin{abstract}
Generalized entropies and relative entropies are the subject of
active research. Similar to the standard relative entropy, the
relative $q$-entropy is generally unbounded for $q>1$. Upper
bounds on the quantum relative $q$-entropy in terms of norm
distances between its arguments are obtained in finite-dimensional
context. These bounds characterize a continuity property in the
sense of Fannes.
\end{abstract}



\maketitle

\section{Introduction}\label{itrn}

The relative entropy has been adopted as very important measure
of statistical distinguishability. In the classical regime, the
relative entropy of probability distribution $\{a_i\}$ to
$\{b_i\}$ is defined by \cite{nielsen} $D(a_i||b_i):=-\sum_i
a_i\ln({b_i}/{a_i})$ and also known as Kullback-Leibler divergence
\cite{KL51}. The usual convention is that $-0\ln0\equiv0$,
$-p_i\ln0\equiv+\infty$ for $p_i>0$. The quantum relative entropy
for (normalized) density operators $\rho$ and $\sigma$ is defined
as \cite{nielsen}
\begin{equation}
\rd(\rho||\sigma):=\tr(\rho\ln\rho-\rho\ln\sigma)
\ . \label{relan}
\end{equation}
This expression is well-defined whenever the kernel of $\sigma$
does not intersect with the support of $\rho$, i.e.
$\kr(\sigma)\subset\kr(\rho)$ \cite{nielsen,ruskai}. Otherwise,
the relative entropy is defined to be $+\infty$. Many fundamental
results of quantum information theory are closely related to
properties of the relative entropy \cite{nielsen,vedral02}.

There exist various generalizations of the standard entropic
functionals. The Tsallis entropy has been found to be very useful
in numerous problems of physics and other sciences \cite{gmt}. In
the context of statistical physics, Tsallis defined the
$q$-entropy for $q\neq1$ by \cite{tsallis}
\begin{equation}
S_q(a_i):=(1-q)^{-1} \left(\sum\nolimits_i a_i^{q} - 1 \right)
\equiv-\sum\nolimits_i a_i^q\ln_q a_i
\end{equation}
where the $q$-logarithm $\ln_q x=\left(x^{1-q}-1\right)/(1-q)$.
This functional has also been derived within an axiomatic
approach \cite{aczel}. Entropic uncertainty principle has been
expressed in terms of the Tsallis entropies \cite{majer,rast104}.
A generalization of the relative entropy for classical probability
distributions is naturally introduced as \cite{borland}
\begin{equation}
D_q(a_i||b_i):=-\sum_i a_i\ln_q(b_i/a_i)\equiv(1-q)^{-1}
\left(1-\sum\nolimits_i a_i^{q}b_i^{1-q}\right)
\ . \label{dqdef}
\end{equation}
Note that the term (\ref{dqdef}) can be recast as the sum
$\sum_ia_i^{q}(\ln_qa_i-\ln_qb_i)$, similar to the standard
analytical form. In the quantum case, the sums in Eq.
(\ref{dqdef}) are replaced by the traces of corresponding
fractional powers of density operators. For $q\in(0;1)$, when no
singularities occur, basic properties of the quantum relative
$q$-entropy are examined in Refs. \cite{abe03,fky04}. Up to a
factor, the relative $q$-entropy is a particular case of Petz's
quasi-entropies \cite{petz86,petz08}.  In the case $q>1$,
singularities may occur in the expression for the relative
$q$-entropy. Like the standard relative entropy, the relative
$q$-entropy is defined to be $+\infty$ in the singular case.

Relations between distinguishability measures are of interest. For
certain applications, some of them may be more appropriate than
others. For instance, cloning processes are often studied with
respect to fidelity-based measures \cite{pati08}, but the relative
entropy has found use as well \cite{rast0971}. Relations of such a
kind are frequently expressed in the form of inequalities. Various
upper bounds on the relative entropy (\ref{relan}) were obtained
in Ref. \cite{AE05}. As given in terms of difference distances,
these bounds characterize a continuity property in the sense of
Fannes \cite{AE05}. Recall that Fannes' inequality bounds from
above a potential change of the von Neumann entropy in terms of
trace norm distance \cite{fannes}. Fannes' inequality has been
extended to the Tsallis $q$-entropy \cite{yanagi,zhang} and its
partial sums \cite{rast1023}. In the classical regime, continuity
properties of wide classes of entropies and relative entropies are
considered in Ref. \cite{naudts04}. In the present paper, we are
interested in upper continuity bounds on the relative $q$-entropy
for $q>1$.

\section{Definitions and background}\label{fdkb}

In this section, the definition of the relative $q$-entropy for
$q>1$ and related questions are discussed. Required mathematical
tools are briefly outlined. By $\kr(\ax)$ we denote the
kernel of operator $\ax$. The support $\sop(\ax)$ is the subspace
orthogonal to $\kr(\ax)$.

\begin{Def}\label{maindf}
For $q>1$ and any pair of normalized density operators $\rho$ and
$\sigma$, the quantum relative $q$-entropy is defined by
\begin{equation}
\rd_{q}(\rho||\sigma):=
\begin{cases}
\frac{1}{1-q}{\,}\Bigl(1-\tr(\rho^{q}\sigma^{1-q})\Bigr)\ ,
& {\ } \kr(\sigma)\subset\kr(\rho) \ , \\
+\infty\ , & {\ } {\rm{otherwise}}\ .
\end{cases}
\label{reendef}
\end{equation}
\end{Def}

Let $\{a\}=\spc(\rho)$ and $\{b\}=\spc(\sigma)$ denote the spectra
of $\rho$ and $\sigma$, and let $\{|a\rangle\}$ and
$\{|b\rangle\}$ denote the related orthonormal bases. Let $f(x)$
and $g(x)$ be functions of scalar variable $x$. Recall that the
formula
\begin{equation}
\tr\bigl(f(\rho){\,}g(\sigma)\bigr)=\sum\nolimits_{a}\sum\nolimits_{b}
|\langle{a}|b\rangle|^2 f(a){\,}g(b)
\label{tradf}
\end{equation}
is regarded as the definition of
$\tr\bigl(f(\rho){\,}g(\sigma)\bigr)$ when $f(\rho)$ or
$g(\sigma)$ is unbounded. The trace in (\ref{reendef}) is written
as
\begin{equation}
\tr(\rho^{q}\sigma^{1-q})=\sum\nolimits_{a}\sum\nolimits_{b}|\langle{a}|b\rangle|^2
a^q b^{1-q} \ , \label{rqsq}
\end{equation}
which is generally unbounded for $q>1$ and singular $\sigma$. If
$\kr(\sigma)\subset\kr(\rho)$, i.e.
$\sop(\rho)\subset\sop(\sigma)$, then we correctly define this
trace as the trace taken just over $\sop(\sigma)$, i.e. the sign
$\sum_b$ is interpreted as $\sum_{b\neq0}$. Effectively, the sum
with respect to $a$ is also restricted to the nonzero $a$'s.

The quantum relative $q$-entropy enjoys some properties similarly
to the standard relative entropy (\ref{relan}). In particular, the
measure (\ref{reendef}) is positive and vanishes only for
$\rho=\sigma$. It is also pseudoadditive for all $q>1$. These
points can be established by a relevant modification of the
reasons given for $q\in(0;1)$ in Ref. \cite{fky04}. One of most
important properties of the standard relative entropy
(\ref{relan}) is its monotonicity under the action of stochastic
maps. The relative $q$-entropy is jointly convex and monotone for
$0\leq{q}\leq2$. These properties follow from the results of Refs.
\cite{naresh09,hmp10}. More details and conditions for equality
can be found in the recent review \cite{JR10}.

Let ${\cal{L}}(\hh)$ be the space of linear operators on
$d$-dimensional Hilbert space $\hh$. By ${\cal{L}}_{+}(\hh)$ and
${\cal{L}}_{++}(\hh)$ we denote the sets of positive and strictly
positive operators respectively. For any $\ax\in{\cal{L}}(\hh)$,
we put $|\ax|=\sqrt{\ax^{\dagger}\ax}\in{\cal{L}}_{+}(\hh)$. The
eigenvalues of $|\ax|$ counted with their multiplicities are the
singular values $s_j(\ax)$ of $\ax$ \cite{bhatia}. For $p\geq1$,
the Schatten $p$-norm of operator $\ax$ is given by
\cite{bhatia,watrous1}
\begin{equation}
\|\ax\|_p=\left(\sum\nolimits_{j=1}^{d} s_j(\ax)^p \right)^{1/p}
\ . \label{schndef}
\end{equation}
This family includes the trace norm $\|\ax\|_{1}$ for $p=1$ and
the spectral norm $\|\ax\|_{\infty}=\max\{s_j(\ax):{\>}1\leq j\leq
d\}$ for $p=\infty$. These norms and relations between them have
found use in various questions of quantum information
\cite{watrous2,rast091,rast103}. For each $p\in[1;\infty]$ and
$\ax,\ay,\az\in{\cal{L}}(\hh)$, there holds (see, e.g., section
2.4 in \cite{watrous1})
\begin{equation}
\|\ax\ay\az\|_p\leq\|\ax\|_{\infty}{\,}\|\ay\|_p{\,}\|\az\|_{\infty}
\ . \label{schnp}
\end{equation}
Since the Schatten $p$-norms are non-increasing in $p$, these norms
satisfy submultiplicativity
$\|\ax\ay\|_p\leq\|\ax\|_p{\>}\|\ay\|_p$. Combining Eq.
(\ref{schnp}) for $p=1$ with $|\tr(\ax)|\leq\tr|\ax|$, we obtain
\begin{equation}
\bigl|\tr(\ax\ay\az)\bigr|\leq\|\ax\|_{\infty}{\,}\|\az\|_{\infty}{\,}\tr|\ay|
\ . \label{xzty}
\end{equation}
We will extensively use the integral representations of matrix
fractional power based on the formulas
\begin{equation}
a^{r}=\frac{\sin r\pi}{\pi} \int\nolimits_0^{\infty}x^{r-1}dx{\>}\frac{a}{a+x}
=\frac{\sin r\pi}{\pi}  \int\nolimits_0^{\infty}y^{-r}dy{\>}\frac{1}{y+a^{-1}}
\qquad (0<r<1) \ . \label{rint}
\end{equation}
The second integral follows from the first one by substituting $x=1/y$.

\section{Upper bounds for $1<q\leq2$}\label{upcb12}

Let us briefly mention lower continuity bounds. Lower bounds on
the relative $q$-entropy for $1<q\leq2$ can be derived from the
inequalities
\begin{equation}
\rd_p(\rho||\sigma)\leq\rd_1(\rho||\sigma)\leq\rd_q(\rho||\sigma)
\ , \label{p1qin}
\end{equation}
where $0\leq p<1$. The relations (\ref{p1qin}) are actually proved
in Ref. \cite{still90}. (Note that the definition of relative
entropy in Ref. \cite{still90} differs from (\ref{relan}) in the
sign, and the formula (\ref{p1qin}) is obtained as the formula (9)
of that paper with the reversed sign.) The quantum relative
entropy obeys (see, e.g., theorem 1.15 in Ref. \cite{ohya})
\begin{equation}
\frac{1}{2}{\>}\|\rho-\sigma\|_1^2\leq \rd_1(\rho||\sigma)
\ . \label{qpins}
\end{equation}
This is a quantum analog of the Pinsker inequality from
classical information theory. Other lower bounds on the relative
entropy (\ref{relan}) are presented in Ref. \cite{AE05}. By Eq.
(\ref{p1qin}), these lower bounds are all valid for
$\rd_q(\rho||\sigma)$ with $1<q\leq2$. Due to this fact and
unboundedness of relative $q$-entropy for $q>1$, we focus on
nontrivial upper bounds. The first upper bound extends the
derivation given for the relative entropy (\ref{relan}) in Ref.
\cite{BR2} (see Example 6.2.31).

\begin{Thm}\label{fup}
If both the density operators $\rho$ and $\sigma$ are strictly
positive then for $1<q\leq2$,
\begin{align}
\rd_q(\rho||\sigma)&\leq \frac{1}{q-1}{\ }\frac{a_1^{q-1}}{\lambda_{0}^{q}}{\>}\|\rho-\sigma\|_{\infty}
\leq\frac{1}{2(q-1)}{\ }\frac{a_1^{q-1}}{\lambda_{0}^{q}}{\>}\|\rho-\sigma\|_{1}
\ , \label{fub1}\\
\rd_q(\rho||\sigma)&\leq \frac{1}{q-1}{\ }\frac{a_1^{q}}{\lambda_{0}^{q}}{\>}\|\rho-\sigma\|_{1}
\ , \label{fub2}
\end{align}
where $a_1:=\max\{a:{\>}a\in\spc(\rho)\}$,
$\lambda_{0}:=\min\bigl\{\lambda:{\>}\lambda\in\spc(\rho)\cup\spc(\sigma)\bigr\}$.
\end{Thm}

{\bf Proof.} Using the second integral from Eq. (\ref{rint}) with
$q-1=r\in(0;1)$ and the linearity of the trace, one gives
\begin{align}
r\rd_q(\rho||\sigma)=\tr\left\{\rho^q(\sigma^{-r}-\rho^{-r})\right\}
&=\tr\left\{\rho^q
{\ }\frac{\sin r\pi}{\pi} \int\nolimits_0^{\infty}y^{-r}dy\left(
(y\pen+\sigma)^{-1}-(y\pen+\rho)^{-1}\right)\right\}
\nonumber\\
&=\frac{\sin r\pi}{\pi} \int\nolimits_0^{\infty}y^{-r}dy
{\>}\tr\Big(\rho^q(y\pen+\sigma)^{-1}\Delta{\>}(y\pen+\rho)^{-1}\Big)
\ . \label{fubin}
\end{align}
Here we used $\am^{-1}-\bm^{-1}=\am^{-1}(\bm-\am)\bm^{-1}$ and
put $\Delta=\rho-\sigma$. Due to Eq. (\ref{xzty}),
the relation (\ref{fubin}) leads to
\begin{equation}
r\rd_q(\rho||\sigma)\leq\tr\bigl|\Delta{\,}\rho^q\bigr|{\ }\frac{\sin r\pi}{\pi} \int\nolimits_0^{\infty}y^{-r}dy
{\,}\left\|(y\pen+\sigma)^{-1}\right\|_{\infty}\left\|(y\pen+\rho)^{-1}\right\|_{\infty}
\leq\tr\bigl|\Delta{\,}\rho^q\bigr|{\ }\frac{1}{\lambda_0^q}
\ . \label{imfub}
\end{equation}
Indeed, the eigenvalues of $(y\pen+\sigma)^{-1}$ are equal to
$(y+b)^{-1}$, $b\in\spc(\sigma)$, and we find
$\left\|(y\pen+\sigma)^{-1}\right\|_{\infty}=(y+b_0)^{-1}$ in
terms of $b_0=\min\{b:{\>}b\in\spc(\sigma)\}$. Putting
$\lambda_0=\min\{a_0,b_0\}$, we further obtain
\begin{equation}
\frac{\sin r\pi}{\pi} \int\nolimits_0^{\infty}y^{-r}dy
{\ }\frac{1}{(y+b_0)(y+a_0)}=\frac{b_0^{-r}-a_0^{-r}}{a_0-b_0}\leq\frac{1}{\lambda_0^q}
\ . \label{inein}
\end{equation}
From Eq. (\ref{schnp}), we get
$\|\ax\ay\|_1\leq\|\ax\|_{\infty}{\,}\|\ay\|_1$. By
$\tr\bigl|\Delta{\,}\rho^q\bigr|\leq\|\Delta\|_{\infty}{\,}\tr(\rho^q)\leq
a_1^{q-1}\|\Delta\|_{\infty}$, we have the first inequality in Eq.
(\ref{fub1}). The second inequality in Eq. (\ref{fub1}) holds due
to the fact that the operator $\Delta$ is traceless and, hence,
$\|\Delta\|_{\infty}\leq(1/2){\,}\|\Delta\|_1$ (see lemma 4 in
Ref. \cite{AE05}). In view of
$\tr\bigl|\Delta{\,}\rho^q\bigr|\leq\|\rho^q\|_{\infty}{\,}\|\Delta\|_1=a_1^q{\,}\|\Delta\|_1$,
the relation (\ref{imfub}) gives Eq. (\ref{fub2}) as well. The
above arguments go through for all $q\in(1;2)$. Since the term
(\ref{rqsq}) is a continuous function of $q$, the claim remains
valid for $q=2$. $\blacksquare$

The bounds (\ref{fub1}) and (\ref{fub2}) are linear in the
corresponding distance between $\rho$ and $\sigma$ and
characterize a continuity of the relative $q$-entropy in the
Fannes sense. The former is more appropriate, when the maximal
eigenvalue $a_1$ of $\rho$ is unknown and replaced with one. The
latter is stronger for sufficiently small values of $a_1$. They
both show that the $\rd_q(\rho||\sigma)$ increases no faster than
$\lambda_{0}^{-q}$ as $\lambda_0$ goes to zero.  However, we would
be interested in a bound which involves only the minimal
eigenvalue of $\sigma$. This can be obtained by a $q$-parametric
extension of the quadratic upper bound from Ref. \cite{AE05} (see
theorem 2 therein). Here one auxiliary statement is required.

\begin{Lem}\label{frcon}
For any two $\am,\bm\in{\cal{L}}_{++}(\hh)$ and $0<r<1$, there holds
\begin{equation}
\am^{-r}-\bm^{-r}\leq\frac{\sin r\pi}{\pi} \int\nolimits_0^{\infty}y^{-r}dy
{\,}(y{\textup{\pen}}+\am)^{-1}(\bm-\am)(y{\textup{\pen}}+\am)^{-1}
\ . \label{aulem}
\end{equation}
\end{Lem}

{\bf Proof.}
The left-hand side of Eq. (\ref{aulem}) can be recast as
$g(\bm)-g(\am)$, where the function $g(t)=-t^{-r}$ is operator
concave for all $r\in(0;1)$. The concavity leads to
\begin{equation}
(1-\theta)g(\am)+\theta g(\bm)\leq g\bigl((1-\theta)\am+\theta\bm\bigr)=g(\am+\theta\dm)
\ , \label{corcon}
\end{equation}
where $\dm=\bm-\am$ and $0<\theta\leq1$. Due to the integral
(\ref{rint}), we rewrite (\ref{corcon}) as
\begin{equation}
g(\bm)-g(\am)\leq\frac{g(\am+\theta\dm)-g(\am)}{\theta}=
\frac{\sin r\pi}{\pi} \int\nolimits_0^{\infty}y^{-r}dy
{\,}(y\pen+\am)^{-1}\dm{\,}(y\pen+\am+\theta\dm)^{-1}
\ .\label{drivt}
\end{equation}
In the limit $\theta\to+0$, the right-hand side of Eq.
(\ref{drivt}) tends to the right-hand side of Eq. (\ref{aulem}),
which is herewith the Fr\'{e}chet differential of $g(\centerdot)$
at $\am$ in the direction $\dm$. $\blacksquare$

\begin{Thm}\label{secup}
For $1<q\leq2$, if $\kr(\sigma)\subset\kr(\rho)$ then
\begin{equation}
\rd_q(\rho||\sigma)\leq
\frac{\ln_q(b_1/b_0)}{1-b_0/b_1}{\ }
\frac{a_1^{q-1}}{b_0^{q-1}}{\>}\|\rho-\sigma\|_1
+\frac{a_1^{q-1}}{b_0^{q}}{\>}\|\rho-\sigma\|_{\infty}{\>}\|\rho-\sigma\|_1
\ , \label{secbn}
\end{equation}
where $a_1:=\max\{a:{\>}a\in\spc(\rho)\}$,
$b_1:=\max\{b:{\>}b\in\spc(\sigma)\}$,
$b_0:=\min\{b\neq0:{\>}b\in\spc(\sigma)\}$.
\end{Thm}

{\bf Proof.} Since $\ax\leq\ay$ gives $\tr(\am\ax)\leq\tr(\am\ay)$
for all $\am\in{\cal{L}}_{+}(\hh)$, the formula (\ref{aulem}) and
the trace linearity lead to
\begin{equation}
\tr\left\{\rho^q(\sigma^{-r}-\rho^{-r})\right\}
\leq\frac{\sin r\pi}{\pi} \int\nolimits_0^{\infty}y^{-r}dy
{\>}\tr\Big(\rho^{q-1}(\sigma+\Delta)(y\pen+\sigma)^{-1}\Delta{\,}(y\pen+\sigma)^{-1}\Big)
\ . \label{subin}
\end{equation}
The right-hand side of Eq. (\ref{subin}) can be rewritten as the
sum of two terms. Using Eq. (\ref{xzty}) and the
submultiplicativity of the spectral norm, the first term is
estimated from above by
\begin{align}
&\frac{\sin r\pi}{\pi} \int\nolimits_0^{\infty}y^{-r}dy
\left|\tr\Big(\rho^{q-1}\sigma{\,}(y\pen+\sigma)^{-1}\Delta{\,}(y\pen+\sigma)^{-1}\Big)\right|
\nonumber\\
&\leq\|\rho^{q-1}\|_{\infty}{\>}\tr|\Delta|{\ }
\frac{\sin r\pi}{\pi} \int\nolimits_0^{\infty}y^{-r}dy
\left\|\sigma(y\pen+\sigma)^{-1}\right\|_{\infty}\left\|(y\pen+\sigma)^{-1}\right\|_{\infty}
\nonumber\\
&=a_1^{q-1}{\,}\|\Delta\|_{1}{\ }
\frac{\sin r\pi}{\pi} \int\nolimits_0^{\infty}y^{-r}dy{\>}\frac{b_1}{y+b_1}{\ }\frac{1}{y+b_0}
=a_1^{q-1}{\,}\|\Delta\|_{1}{\,}b_0^{-r}{\>}\frac{1-(b_0/b_1)^r}{1-b_0/b_1}
\ . \label{subiw1}
\end{align}
Here we have inserted
$\left\|\sigma(y\pen+\sigma)^{-1}\right\|_{\infty}=b_1(y+b_1)^{-1}$,
since the eigenvalues of $\sigma(y\pen+\sigma)^{-1}$ are equal to
$b(y+b)^{-1}$, and the latter is an increasing function of $b$.
Dividing the right-hand side of Eq. (\ref{subiw1}) by $r=q-1$, we
obtain the first term of the right-hand side of Eq. (\ref{secbn}).
Using Eq. (\ref{xzty}) and the submultiplicativity again, the
second term of the right-hand side of Eq. (\ref{subin}) is
estimated by
\begin{align}
&\frac{\sin r\pi}{\pi} \int\nolimits_0^{\infty}y^{-r}dy
\left|\tr\Big(\rho^{q-1}\Delta{\,}(y\pen+\sigma)^{-1}\Delta{\,}(y\pen+\sigma)^{-1}\Big)\right|
\nonumber\\
&\leq\|\rho^{q-1}\|_{\infty}{\>}\|\Delta\|_{\infty}{\>}\tr|\Delta|{\ }
\frac{\sin r\pi}{\pi} \int\nolimits_0^{\infty}y^{-r}dy
\left\|(y\pen+\sigma)^{-1}\right\|_{\infty}^2
\nonumber\\
&=a_1^{q-1}{\,}\|\Delta\|_{\infty}{\>}\|\Delta\|_{1}{\ }
\frac{\sin r\pi}{\pi} \int\nolimits_0^{\infty}y^{-r}dy{\>}\frac{1}{(y+b_0)^2}
=a_1^{q-1}{\,}\|\Delta\|_{\infty}{\>}\|\Delta\|_{1}{\,}rb_0^{-r-1}
\ . \label{subiw2}
\end{align}
Dividing this by $r=q-1$, we obtain the second term of the
right-hand side of Eq. (\ref{secbn}). $\blacksquare$

The bound (\ref{secbn}) is not purely quadratic, but its second
term dominant for small $b_0$ is just quadratic in a distance
between $\rho$ and $\sigma$. For traceless $\Delta=\rho-\sigma$,
we have $\|\Delta\|_{\infty}\leq(1/2){\,}\|\Delta\|_1$ (see lemma
4 in Ref. \cite{AE05}), whence
\begin{equation}
\rd_q(\rho||\sigma)\leq
\frac{\ln_q(b_1/b_0)}{1-b_0/b_1}{\ }
\frac{a_1^{q-1}}{b_0^{q-1}}{\>}\|\rho-\sigma\|_{1}
+\frac{a_1^{q-1}}{2b_0^{q}}{\>}\|\rho-\sigma\|_{1}^2
\ . \label{secbn22}
\end{equation}
The upper continuity bounds (\ref{secbn}) and (\ref{secbn22})
involve the minimal eigenvalue only of $\sigma$ and suit for
singular $\rho$. These points are advantages of these bounds. At
the same time, we would like to find an upper bound with the
dependence $b_0^{1-q}$ in view of the term $\sigma^{1-q}$ in the
definition of $\rd_q(\rho||\sigma)$. Indeed, for the standard
relative entropy (\ref{relan}) we have the bound which is
logarithmic in the minimal eigenvalue of $\sigma$ \cite{AE05}.

\section{Upper bounds for arbitrary $q>1$}\label{upcb21}

In this section, we present upper continuity bounds that cover all
the values $q>1$. Moreover, these bounds has the dominant term
with a dependence $b_0^{1-q}$ in the minimal eigenvalue of
$\sigma$. We first consider the case of integer powers.

\begin{Lem}\label{sbmn}
For integer $n\geq1$ and $\ax,\ay\in{\cal{L}}(\hh)$, there holds
\begin{equation}
\|\ax^n-\ay^n\|_{p}\leq{n}\lambda_{1}^{n-1}\|\ax-\ay\|_{p}
\ , \label{lemsm}
\end{equation}
where $\lambda_{1}:=\max\bigl\{\|\ax\|_{\infty},\|\ay\|_{\infty}\bigr\}$.
\end{Lem}

{\bf Proof.} We shall proceed by induction. For $n=1$, the claim
is obvious. For $n>1$, we write
\begin{equation}
\ax^{n+1}-\ay^{n+1}=\ax^{n+1}-\ax^n\ay+\ax^n\ay-\ay^{n+1}=
\ax^n(\ax-\ay)+(\ax^n-\ay^n)\ay
\ . \label{axnyn}
\end{equation}
By means of the triangle inequality, Eq. (\ref{schnp}) and
$\|\ax^n\|_{\infty}\leq\lambda_{1}^{n}$, we then obtain
\begin{equation}
\|\ax^{n+1}-\ay^{n+1}\|_{p}\leq\|\ax^n(\ax-\ay)\|_{p}+\|(\ax^n-\ay^n)\ay\|_{p}
\leq\lambda_{1}^{n}\|\ax-\ay\|_{p}+\lambda_{1}\|\ax^n-\ay^n\|_{p}
\ . \label{axnyn1}
\end{equation}
Assuming Eq. (\ref{lemsm}), we derive
$\|\ax^{n+1}-\ay^{n+1}\|_{p}\leq(n+1)\lambda_{1}^{n}\|\ax-\ay\|_{p}$ too.
$\blacksquare$

\begin{Rem}\label{subbnm}
Let $\|\centerdot\|$ be a submultiplicative norm and
$\ax,\ay\in{\cal{L}}(\hh)$. For integer $n\geq1$,  there holds
\begin{equation}
\|\ax^n-\ay^n\|\leq{n}{\,}\theta^{n-1}\|\ax-\ay\|
\ . \label{remsm}
\end{equation}
where $\theta:=\max\bigl\{\|\ax\|,\|\ay\|\bigr\}$. The claim can be
proved similarly to the statement of Lemma \ref{sbmn}.
\end{Rem}

Using Lemma \ref{sbmn}, we now obtain the upper bound of a kind
$b_0^{1-q}$ for any $q>1$, including non-integer $q$. By
$\lfloor{q}\rfloor$ and $\lceil{q}\rceil$ we respectively denote
the floor and ceiling of real $q$. The main result of this section
is stated as follows.

\begin{Thm}\label{b0q1}
For arbitrary $q>1$, if $\kr(\sigma)\subset\kr(\rho)$ then
\begin{equation}
\rd_q(\rho||\sigma)\leq\frac{\lceil{q}\rceil-1}{q-1}{\ }
\frac{\lambda_{1}^{q-1}}{b_{0}^{q-1}}
{\>}\|\rho-\sigma\|_{1}
\ , \label{dqm1}
\end{equation}
where
$\lambda_{1}:=\max\bigl\{\lambda:{\>}\lambda\in\spc(\rho)\cup\spc(\sigma)\bigr\}$,
$b_0:=\min\{b\neq0:{\>}b\in\spc(\sigma)\}$.
\end{Thm}

{\bf Proof.} For non-integer $q>1$, we write $n=\lfloor{q}\rfloor$,
$s=q-n$, and the identity
\begin{equation}
\rho^{n+s}-\sigma^{n+s}=(\rho^n-\sigma^n)\rho^s+\sigma^n(\rho^s-\sigma^s)
\ . \label{rsns}
\end{equation}
By the linearity of the trace and $r=q-1$, we then have
\begin{equation}
r\rd_q(\rho||\sigma)=\tr\left\{\sigma^{-r}(\rho^q-\sigma^q)\right\}
=\tr\left\{\sigma^{-r}(\rho^n-\sigma^n)\rho^s\right\}+\tr\left\{\sigma^{1-s}(\rho^s-\sigma^s)\right\}
\ . \label{rsns1}
\end{equation}
Due to Eqs. (\ref{xzty}) and (\ref{lemsm}), the first trace in the
right-hand side of Eq. (\ref{rsns1}) is bounded from above by the
quantity
\begin{equation}
\|\sigma^{-r}\|_{\infty}{\,}\|\rho^n-\sigma^n\|_1{\,}\|\rho^{s}\|_{\infty}
\leq{n}{\,}a_1^s\lambda_{1}^{n-1}b_0^{-r}{\,}\|\rho-\sigma\|_1
\leq\lfloor{q}\rfloor\lambda_{1}^{q-1}b_0^{-r}{\,}\|\rho-\sigma\|_1
\ . \label{rsns2}
\end{equation}
The second trace is merely
$\tr\big(\sigma^{1-s}\rho^s\big)-1=-(1-s)\rd_s(\rho||\sigma)\leq0$
in view of $s\in(0;1)$ and the positivity of the relative
$s$-entropy. Combining this fact with Eqs. (\ref{rsns1}) and
(\ref{rsns2}) then gives the claim for non-integer $q>1$. For
$q\in(n;n+1)$, the right-hand side of Eq. (\ref{rsns2}) is a
continuous function of the parameter $q$. The expression for
$\rd_q(\rho||\sigma)$ is continuous for all $q\neq1$. So the bound
(\ref{dqm1}) remains valid when $q$ tends to an integer from
below. $\blacksquare$

With respect to small $b_0$, the right-hand side of Eq.
(\ref{dqm1}) has a dependence $b_0^{1-q}$. So the upper continuity
bound (\ref{dqm1}) is stronger than the bounds of Theorems
\ref{fup} and \ref{secup}. The upper bound (\ref{dqm1}) also
involves $\lambda_{1}^{q-1}$ and the trace norm distance between
density operators. By a structure, the right-hand side of Eq.
(\ref{dqm1}) is in good agreement with the expression for the
relative $q$-entropy. Another advantage of the bound (\ref{dqm1})
is that it covers the range $q>2$ as well. At the same time, the
factor $\left(\lceil{q}\rceil-1\right)/(q-1)$ is not continuous in
$q$. For very interesting values, $1<q\leq2$, the relation
(\ref{rsns2}) leads to the inequality
\begin{equation}
\rd_q(\rho||\sigma)\leq
\frac{1}{q-1}{\ }\frac{a_{1}^{q-1}}{b_{0}^{q-1}}{\ }\|\rho-\sigma\|_1
\ , \label{dqm2}
\end{equation}
with $a_{1}$ instead of $\lambda_{1}$. In general, the upper
continuity bound (\ref{dqm2}) seems to be more appropriate among
the ones presented above for $1<q\leq2$. For sufficiently small
$b_{0}$, this bound is clearly tighter than (\ref{secbn}) and
(\ref{secbn22}). On the other hand, the upper bounds (\ref{secbn})
and (\ref{secbn22}) may be more sharpening in the sense of
closeness of the states $\rho$ and $\sigma$, when $b_{0}$ is not
very small. Due to $a_{1}\leq\lambda_{1}\leq1$, all the above
bounds can be rewritten with one instead of the maximal
eigenvalue. Finally, we note that our methods could be used for
estimating the modulus of the second trace in the right-hand side
of Eq. (\ref{rsns1}). We present one continuity bound of such a
kind, though it was not required for the proof of Theorem
\ref{b0q1}.

\begin{Lem}\label{abst}
Let $\am\in{\cal{L}}_{+}(\hh)$, $\bm\in{\cal{L}}_{++}(\hh)$ and
$\tr(\am)=\tr(\bm)=\tau$. For $0<s<1$, there holds
\begin{equation}
\left|\tr\big(\bm^{1-s}\am^{s}\big)-\tau\right|\leq
\frac{a_{1}^{s}}{b_{0}^{s}}{\>}\|\am-\bm\|_{1}
\ , \label{ab10}
\end{equation}
where $a_{1}:=\max\bigl\{a:{\>}a\in\spc(\am)\bigr\}$,
$b_{0}:=\min\bigl\{b:{\>}b\in\spc(\bm)\bigr\}$.
\end{Lem}

{\bf Proof.} Using the first integral representation from Eq.
(\ref{rint}) and the properties of the trace, we have
\begin{align}
&\tr\big(\bm^{-s}\am^{s}{\,}\bm\big)=\frac{\sin s\pi}{\pi} \int\nolimits_0^{\infty}x^{s-1}dx
{\>}\tr\Big(\bm^{-s}\am(\am+x\pen)^{-1}\bm\Big)
\ , \label{bas}\\
&\tau=\tr\big(\bm^{-s}\am{\,}\bm^{s}\big)=\frac{\sin s\pi}{\pi} \int\nolimits_0^{\infty}x^{s-1}dx
{\>}\tr\Big(\bm^{-s}\am(\bm+x\pen)^{-1}\bm\Big)
\ . \label{sab}
\end{align}
Combining Eqs. (\ref{bas}) and (\ref{sab}) with the identity
\begin{equation}
\am(\am+x\pen)^{-1}\bm-\am(\bm+x\pen)^{-1}\bm=
\am(\am+x\pen)^{-1}\bigl\{(\bm+x\pen)-(\am+x\pen)\bigr\}(\bm+x\pen)^{-1}\bm
\label{idnab}
\end{equation}
and putting $\dm=\bm-\am$, we obtain
\begin{equation}
\tr\big(\bm^{1-s}\am^{s}\big)-\tau=\frac{\sin s\pi}{\pi} \int\nolimits_0^{\infty}x^{s-1}dx
{\>}\tr\Big(\bm^{-s}\am(\am+x\pen)^{-1}\dm{\,}(\bm+x\pen)^{-1}\bm\Big)
\ . \label{abd1}
\end{equation}
Using the cyclic property of the trace and Eq. (\ref{xzty}), one
gets
\begin{equation}
\left|\tr\big(\bm^{1-s}\am^{s}\big)-\tau\right|
\leq\tr|\dm|{\ }\frac{\sin s\pi}{\pi}\int\nolimits_0^{\infty}x^{s-1}dx\left\|\am(\am+x\pen)^{-1}\right\|_{\infty}
\left\|(\bm+x\pen)^{-1}\bm^{1-s}\right\|_{\infty}
\ . \label{abd2}
\end{equation}
The eigenvalues of $\am(\am+x\pen)^{-1}$ are equal to
$a(a+x)^{-1}\leq{a}_1(a_1+x)^{-1}$, the eigenvalues of
$(\bm+x\pen)^{-1}\bm^{1-s}$ are equal to
$(b+x)^{-1}b^{1-s}\leq{b}^{-s}\leq{b}_0^{-s}$, whence
\begin{equation}
\left\|\am(\am+x\pen)^{-1}\right\|_{\infty}=a_1(a_1+x)^{-1} \ ,
\qquad \left\|(\bm+x\pen)^{-1}\bm^{1-s}\right\|_{\infty}\leq{b}_0^{-s}
\ . \label{aa1}
\end{equation}
Due to these relations and $\tr|\dm|=\|\am-\bm\|_{1}$, the
inequality (\ref{abd2}) leads to
\begin{equation}
\left|\tr\big(\bm^{1-s}\am^{s}\big)-\tau\right|\leq\frac{1}{b_0^{s}}{\>}\|\am-\bm\|_{1}
{\>}\frac{\sin s\pi}{\pi} \int\nolimits_0^{\infty}x^{s-1}dx{\>}\frac{a_1}{a_1+x}
\ . \label{ab11}
\end{equation}
By the first integral of Eq. (\ref{rint}), this is equivalent to
Eq. (\ref{ab10}). $\blacksquare$

\section{Summary}\label{sump}

Similar to the standard relative entropy, the Tsallis relative
$q$-entropy is generally unbounded for $q>1$. Hence upper bounds
on this functional are of interest. In this paper, we have
obtained several upper bounds on the relative $q$-entropy for
$q>1$. These bounds are expressed in terms of norm distances
between its two arguments and the minimal eigenvalue of the second
argument. The presented inequalities characterize the property of
continuity of the relative $q$-entropy for states which are close
in the trace norm sense. They also estimate from above the rate of
divergence of the relative $q$-entropy when the minimal eigenvalue
of its second argument goes to zero. The considered upper bounds
can be regarded as some $q$-parametric extensions of those bounds
that have been obtained in the literature for the standard
relative entropy. To derive the results, we have extensively used
general properties of the trace and spectral norms as well as the
known integral representations of matrix fractional powers.

\acknowledgments
The present author is grateful to an anonymous referee for useful comments.

\end{document}